\documentclass{article}

\usepackage{arxiv}

\usepackage[utf8]{inputenc} 
\usepackage[T1]{fontenc}    
\usepackage{hyperref}       
\usepackage{url}            
\usepackage{booktabs}       
\usepackage{amsfonts}       
\usepackage{nicefrac}       
\usepackage{microtype}      
\usepackage{lipsum}		
\usepackage{graphicx}
\usepackage{doi}
\usepackage{wrapfig}

\title{``What artists want'': Elicitation of artist requirements to feed the design on a new collaboration platform for creative work}

\author{
Angeliki Antoniou \\
  Department of Archival, Library \& Information Studies\\
  University of West Attica\\
  \texttt{angelant@uniwa.gr} \\
  \And
Ioanna Lykourentzou \\
  Department of Information and Computing Sciences\\
  Utrecht University\\
  \texttt{i.lykourentzou@uu.nl} \\
  \AND
Antonios Liapis \\
  Institute of Digital Games\\
  University of Malta\\
  \texttt{antonios.liapis@um.edu.mt} \\
  \And
Dimitra Nikolou \\
  ARTWORKS \\
  \texttt{dimitra@art-works.gr} \\
  \And
Marily Konstantinopoulou \\
  ARTWORKS \\
  \texttt{marily@art-works.gr}
}

\begin{document}
\maketitle

\begin{abstract}
Aiming at designing a decentralized platform to support grassroot initiatives for self-organized creative work, the present work solicited feedback from a group of visual artists regarding their work processes and concerns. The paper presents the qualitative methodology followed for collecting requirements from the target audience of the envisioned software solution. The data gathered from the focus group is analyzed and we conclude with a set of important requirements that the future platform needs to fulfill. 
\end{abstract}


\section{Introduction}\label{sec:introduction}
What do artists want? Which concerns and problems during an artist’s workflow could be mitigated through a software platform for group collaboration? Can algorithms intervene in the creative process and is this important for artists, science and societies? 

Attempting to provide answers to the above questions, we started designing the ERICA (Decentralized gRassroot Initiatives for self-organized CreAtive work) concept. We envision ERICA as a decentralized platform to support grassroot initiatives for self-organized creative work. The reason behind this initiative is that the nature of work is evolving to new forms, often assisted by technology. Creative work is no exception. The ongoing Covid19 pandemic seems to be speeding up these changes in workflows, with more and more employers and employees turning to online work. In addition, new and flexible ways of work such as part-time and crowd-work have become mainstream. Different platforms are created to assist these new forms of work, such as Fiverr, Upwork, etc. which use algorithms to allocate workers to tasks, form teams and create processes. Although these algorithms are increasing efficiency, however, they can also be restrictive in the way they overrun human initiatives such as finding collaborators and experimenting with tasks previously unknown to them. We hypothesize that in the long run, such top-down algorithms can hinder creative processes with clear implications at the personal but also the societal level. In this light, ERICA wishes to support bottom-up efforts, to allow creative workers’ freedom of choice, to assist collaboration but also to enhance experimentation with unknown elements (e.g. collaboration between unexpected parties such as artists and medical doctors). 

In the present work we focus on one type of creative workers, the artists; we wish to understand their needs for effective work, collaboration, inspiration, etc. Returning to the central issue of what artists want, in June 2021 we organized a focus group with visual artists and discussed relevant issues. The current work presents the focus group methodology and results, and concludes with elements that ERICA should be implementing.
 
\section{Challenges of On-Demand Work}\label{sec:litreview}
As discussed in Section \ref{sec:introduction}, on-demand work is becoming increasingly common, both offline and online. Crowdsourcing platforms have given both clients and workers access to a marketplace with significant potential especially in creative work. These same platforms however emphasize top-down supervision and client control \cite{kellog2020}. They rely on external reinforcement, reduce worker autonomy, discourage collaboration and distrust worker initiative \cite{lykourentzou2021,retelny2017}. Workers get paid by the product (``piecemeal'' work), often unfairly, and they do not receive any compensation for training, waiting times for the task to start, and so on \cite{alkhatib2017}. The nature of online platforms renders the work of online on-demand workers invisible; platform customers get the impression that they are fine-tuning a platform service when uploading their task, rather than dealing with real people \cite{gray2019}. Similarly, platforms do not allow for creative expression, since they use very rigid workflows that decide exactly how the work will be done \cite{retelny2017}.

\begin{wrapfigure}[39]{r}{0.63\textwidth}
    \centering
    \includegraphics[width=\linewidth]{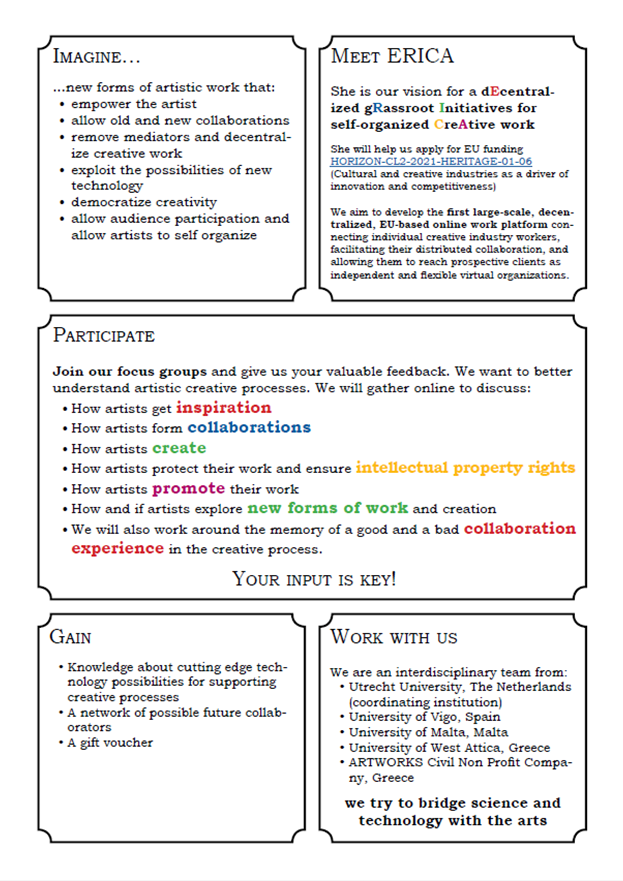}
    \caption{Invitation sent to artists}
    \label{fig:invitation}
\end{wrapfigure}

Moreover, when AI is leveraged to assist in allocating worker resources such as splitting and assigning tasks \cite{roy2015,difallah2019} the worker can not propose changes to this estimation model. As a result, AI is taking an increasingly central role in large-scale online work management, orchestrating people, keeping track of their performance and deciding precisely where, when and with whom each person will work, and how. The more prescriptive an algorithm becomes, the more trust it requires and more ethical issues it raises \cite{martin2019}. Since AI is non-transparent, algorighmic  biases can unilaterally penalize certain workers (e.g. not reaching a certain productivity level can lead to less customers seeing your profile, etc.) \cite{huang2013}.

\section{Collecting Artists' Feedback}\label{sec:method}

In order to collect the creative workers’ concerns and requirements regarding the past and future forms of collaboration, we conducted a user study with visual artists in Greece. Participation in this user study was solicited by ARTWORKS, a Greek nonprofit organization that aims to create a fertile and nurturing environment for Greek artists through funding and public engagement opportunities. ARTWORKS invited visual artists from their network of collaborating artists in order to participate in the study. Figure \ref{fig:invitation} shows the invitation sent to artists, describing the concept. 

Five visual artists participated in this first focus group: two identified as female and three identified as male. Most of the participants knew of each other, but they had never collaborated in the past. From the research team, two members from ARTWORKS were present to ensure that the artists saw a familiar face and felt more comfortable. However, the ARTWORKS members made the necessary introductions at the beginning and remained as audience. Two researchers had an active part, one in coordinating the focus groups overall and one participating in the phase where technology had to be presented (phase 2; see below). Due to health concerns, the group discussion took place virtually, in a Zoom call. The entirety of the process took place in Greek, with the recording being transcribed by the first and second author and translated in English. The focus group session lasted three  hours and was split into three phases. Phase 1 solicited a discussion among artists regarding good and bad work experiences they had had. Then a short break of 15 minutes followed and phase 2 started, where one of the researchers presented technology possibilities and the general idea behind ERICA. During phase 3 artists were asked to consider what technology could offer them and shape future scenarios of use.

\section{Findings \& Discussion}\label{sec:results}

\subsection*{Phase 1}

Artists were first asked to recall a bad work experience they had and share it with the group. Many issues emerged: among the most prominent was that of the  \emph{artist payments and budget} issues. Artists seem not only to struggle with low budgets from exhibition centers and institutions, but also with sudden and unexpected budget cuts; agreements seem to be rather vague, often binding the artists but not the organizers (institutions, venues, etc).  As participant K mentioned: “the project was going very well, but it had a very low budget…we started working…budget was cut (further) to organize a party. How can we refuse? We do not want to lose the community”.  Participant M also commented that “many times they tell us that if you want something more, use money from your payment” and participant K added “many times institutions consider paying us as out of this world…why should they pay you since they will show your work? It is considered normal not to have a payment”. 

The demands for free work are often coupled with issues of artist \emph{recognition} and \emph{intellectual property rights}. Participant M explains that it is often very difficult to share the same understanding with venues and have clear expectations: “After they managed to reach an agreement with a lot of effort, the festival reaped everything and presented them as its own work. Disappointment”. This experience was also common with participant P who said: “they (the curators) kept skyping them for their work, they asked for material, but without having confirmed their participation… although the curators are not usually asking you 3 times for your reports. In the end, they appropriated their reports and wrote in the list of the report whatever they wanted.”

As it seems, \emph{transparency} of processes and decisions is a challenge. Participant G mentioned: “In collective exhibitions, artists often see different budgets and feel a strong hierarchy. Some got paid for a hotel, some did not. Obviously there may be a different budget but there must be transparency.” Participant A added: “The event usually has its own agenda and its own target audience. This agenda is not always obvious. A hierarchy can be seen through the setting or selection of the artists”. But transparency on its own does not seem satisfactory, since injustice is also a part of the picture. “The son of a very large collector participated in an exhibition. He had many benefits. There was transparency there. Class and hierarchical issues arise. Transparency alone is not enough” (participant P) and \emph{fairness} is also necessary. The art institutions often impose such hierarchies, which in turn shape the art market; these hierarchies seem to “create problems in the relationships of artists. This is counterproductive and counter cooperative” (Participant A). 

Could all these factors lead to reduced creativity and loss of talent? This seems to be the case, since even the domain of the arts (which should in principle be one of the freer expressions in societies) is often restricted and seems to follow strict and often shady market rules, or as participant K said: “the system that supports art is the opposite of art itself”. \emph{Freedom of creative expression} is obstructed by curator demands and inflexible processes. Participant A explained: “while they know that I make modifications and installations in the space, they complain to me that I disturb the space. So even though they knew my job, they did not respect me. It seems that our works go through criticism or filters of the curators for what they imagine your work should be. But the brief must be made, that is, to give requirements, but until intervention there is a delicate balance. How can you maintain your values and at the same time balance with the desires of the organization?”

Finally, inflexible processes and \emph{bureaucratic issues} also reduce productivity and creativity, sometimes leading to the cancellation of plans. Participant K shared his experience: “in a collective they had to implement a project. The bureaucracy did not allow the implementation. Initially there were delays, with great time pressures, with demands (on the artists) to put the money on their own and get refunds later. The contract had clauses and the civil liability passed to the artists and not to the municipality. But the municipality wanted its logo everywhere. So it never happened”. 

After all artists shared and discussed their negative experiences, we proceeded with soliciting the discussion of positive experiences. As evidenced from the negative experiences artists had, trust is a catalyst in a positive experience. Trusting the institutions but also other artists during interaction and collaboration is essential to allow artists to work freely. \emph{Collectives}, unions of artists, seem to provide the necessary safety and trust. Participant G mentioned: “we want a relationship of trust. It's ugly to be alone in so many levels. A team helps you to be yourself”. Making these artist groups was mentioned by most participants, as it is a very important element in their work and lives. Collectives seem to have a long duration and within them, artists form collaborations, exchange views, evolve and even make long-lasting friendships. “Collectivity is something positive in the long run. Because there are difficult things, when I'm in a collective; it is really psychotherapeutic. You endure hardships because you have this network. It is valuable. How could this be provided? Are there instructions or directions to make a good collective? That is, to somehow set up a good team?” (participant K). 

It also feels safer for artists to trust \emph{smaller venues} and institutions where personal relations are central. ``Working with smaller spaces and small structures, I have good experiences. The ambience of the institution matters. In smaller spaces you have substance. Large spaces give another prestige to your CV, but in small spaces you feel safe. It is a good practice to work with both small and large spaces. Some people have to cover your back, so that you do not feel unprotected. Insecurity does not allow you to reach your goal'' (Participant P).  Small and clearly defined structures, while they might lack prestige, seem to assist creative expression. 

In addition, it is important for artists to be connected to other artists, events and venues but also art associations that provide some financial safety, such as ARTWORKS. Apart from the collectives, artists need to enrich their network of connections: “Collaborations are also very important. The \emph{connectivity} and financial security that ARTWORKS has given us is very important” (participant M).

Finally, one very important aspect was brought up by participant A: “artists try everything, but they never become experts in something. Testing is the key, but not in depth; you look at everything at once and nothing in depth. This helps you to think outside the box”. The need for multidimensional, even unexpected, experiences is expressed here as a way to feed creativity.  

\subsection*{Phase 2}

After a short break, a researcher with expertise in collaborative, crowd work presented future technological possibilities and what ERICA could offer. ERICA could allow new forms of creative work to emerge by allowing for bottom-up team formation and ideation through technology. It offers possibilities of technology-enhanced creativity such as a theatre play where one of the actors is replaced by artificial intelligence that improvises, or paintings that can sing opera, etc. ERICA would further empower artists by removing mediators and decentralizing creative work. Furthermore, it would allow collaborations to form and old ones to be maintained, permitting artists to self-organize and create independent and flexible virtual organizations. Finally, ERICA would democratize the creative processes with active audience participation and the collaboration between unexpected parties.

\subsection*{Phase 3}

After the presentation of the concept of ERICA and its potential, the artists had to discuss the following topics, taking technological possibilities into account: artistic inspiration, collaborations, matching creative workers, intellectual property rights and ownership, alternative labor models and governance, evaluation of artistic and creative work. 

Regarding \emph{inspiration}, artists favor research and unexpected sources of information. Participant M explained that they get inspiration from research in libraries and academic repositories such as \url{https://www.academia.edu/}. They believe that technology could offer rich information to artists. However, personalization might be an issue, since the algorithms will pre-decide on what different people see. “But if, as on YouTube, which offers the same to similar users, we do not want everyone to get the same. We should not end up in a pool where more or less everyone will receive the same information, because this will be a big problem in artistic creation”. Diversity and uniqueness are therefore very important. In fact, it was also suggested that by changing the roles that creative workers used to take, new unexpected innovation can occur. For example, participant K described a project where scientists can be residents in a creative hub or artists could spend a few months in a scientific laboratory (\url{https://www.studiotopia.eu/}). Thus an important requirement for ERICA is that teams should not lock people in information bubbles, reducing them to a limited set of profile characteristics. ERICA should also allow the flexible matching of people in teams and help bring together people of different profiles, such as artists and scientists, etc. 

Technology could further enhance \emph{collaborations}, since their importance was widely accepted by the participants. Artists actively look for like-minded people with similar interests and even hobbies. Similar interests seem to motivate people to invest time in order to build the collaboration while also giving a sense of common identity. Participant A explained how they actively look for people with similar interests, for example. Sometimes collaborations emerge from an existing network of friends and acquaintances; in other cases complete strangers might end up working together. Participant M said: ``I collaborate with strangers. I invest months to make sure we feel safe. We only need safety and time to understand each other''. It seems that before the creative process can start, a certain amount of trust must be built which needs time to evolve. Participants also mentioned that unexpected collaborations can be very fruitful, since they go beyond possible initial expectations of failure and surprise positively. Once asked how they expand their network with previously unknown individuals, artists explained that it is a trial and error process. In particular, participant P admitted that ``CVs do not help much in collaborations. I do not want such types of filters. With trial and error, I take small steps with smaller initiatives''. Thus, ERICA should incorporate scaffolding elements in its design, allowing creative workers to build collaborations with unknown others, starting from small jobs and escalating to larger ones, once trust is gained, provided that no strict time restrictions apply. Artists also expect failure and believe it is a part of their job, since it becomes a part of self-evaluation and introspection, but also a possibility for a new start. Participant G explained: ``Sometimes it may be fruitful for something to fail, because it is good, maybe for something else. Good cooperation can come from where you do not expect it''. 

However, the question remains on how ERICA can \emph{match creative workers}. Participant A described a project they are involved in: ``(It is an) application for archiving material relevant to space… a hardware repository. They are already making such an application. The application is aimed at creating routes with gamification elements. They seek the help of developers. How to pass the idea and create interest for someone to participate wholeheartedly in the project you tell them? How to create a two-way relationship? We want to create a profile for a person that wants to create his/her identity. Maybe with a concept tag or pool of images, to create a pool of interest…We can be inspired by gaming and gaming communities. How the player becomes part of the development. He/she wants to see how contemporary art can be created through a platform. The key is the connection”. Therefore, ERICA should further focus on motivation schemes and application of participatory design processes \cite{bannon2018}. Since, as explained above, a CV is not by itself adequate to match people, other approaches need to be explored such as sharing of similar values and beliefs, similar aesthetics, without however excluding the unexpected to arise by also allowing very different people to get to know each other. Although the participant believes that ``connection is the key'', ERICA needs to work around delicate balances: on the one hand bringing like-minded people together, on the other ensuring that new and unexpected connections happen. 

Another significant element during collaborations is \emph{ownership and intellectual property rights}. Artists admit that in this creative sector these issues are still unsolved. Participant P said: “who has the ownership? How much exactly? IPRs in practices are not solved at all…there is much that needs to be done. I have all my work open but I always ask for a reference to the creator”. Participant K added: “I use tools from political science to understand what is happening with the issue of creation and ownership. The ownership status on the platform is very important, e.g. are we becoming shareholders? Are we equal? Using models of Co-ownership, for example? Let's look at ethical bank models. Co-ownership is one solution”. Participant A agreed and explained how ERICA should consider different models of co-ownership, since “they will change the way we see things”. On the same topic, participant G explained the differences in the creative process in a physical and a digital environment: “at the carpenter (that you might need to complete your work) you see what is happening. In programming and algorithms you do not know what is happening. It is more hidden. You do not know under the hood the mechanisms of what the program presents to you. The digital environment has clear frameworks and creates a freedom. The problem is in the social structure that we all have to say what we have done. If we have to divide the project and give IPR to each one, we lose the whole. Something could belong to everyone \ldots so we have to look at other ownership models”. Participant M also added: “we need to look at ways to participate. Do you want to do a job yourself or do you want to join a team? These are different ways”. Since all artists agreed that IPR is not straightforward to solve, ERICA needs to explore in-depth issues of ownership and IPR. In addition, some artists were in favor of all team members co-owning from the beginning the possible creative outcome. Thus, ERICA could examine new models of co-ownership and how these affect not only the performance, but also the growth mentality of people. Moreover, ERICA should also consider alternative \emph{labor and governance models}, implying that creative workers can choose how they participate, for example if they join the online collaboration as an equal creator, as a commissioned part or as an apprentice. Technology can facilitate these new ownership and labor models, and facilitate for example more fair payments. However, technology changes the nature of the (creative) work and new governance issues emerge. For example, the rules of behavior within the platform need to be transparent, since different people have different motives, drives and work ethics. 

Finally, artists discussed issues of \emph{evaluation} of artistic and creative work. Participant A said that ``I consider a work to be a good one if I have exhausted researching all viewpoints. When it has reached maturity. When I have approached it from multiple perspectives''. Participant G added that they are ``interested in the process. The outcome is irrelevant sometimes. The process is evaluated. If you focus only on the outcome you lose the open and the ongoing. The end is relevant''. Participant M agreed saying that ``the process needs to be creative. If we only focus on the outcome then this becomes standardized and our work should not be standardized”. But participant A noticed that ``be careful so that the evaluation of the process does not also become standardized''. It seems that artists try to continuously question their work (process and outcome) but it is customary that they ask other artists to evaluate it. Peer review seems important. This is why artists engage in studio visits: to experience, evaluate and discuss works. However, participant K said that ``I say (to the visiting artists) how much criticism I allow every time. If I am feeling good one day I allow more. If I do not feel good, I allow less''. So who evaluates and the level of evaluation seems to be also regulated. Furthermore, evaluation is a continuous process and an incremental approach seems logical. In this light, a question that ERICA could further explore is whether the platform could help creative workers evaluate and also sell the work-in-progress. This could follow the example of other creative industries, such as games that sell early-access versions, downloadable content \cite{tyni2011dlc}, etc.

\section{Conclusions}\label{sec:conclusion}
Wishing to design a decentralized platform to support grassroot initiatives for self-organized creative work, we had to focus on the needs of creative workers. The present study focused on one type of creative workers --visual artists-- in order to collect their requirements. Future work will also collect and analyze information from other creative workers, such as game designers and developers. Implementing a qualitative methodology for the extraction of artists’ requirements, we were able to identify a number of issues that ERICA needs to address. 

First of all, it is important to provide artists with diverse information, different to various people in a way that not everyone is presented with the same content. However, in doing so, one needs to be careful not to provide too much irrelevant information that might be destructive for the creative process. Thus ERICA should find the balance in providing the right amount of information and the right amount of diversity in the information. In addition, it is essential for artists to be able to maintain collaborations and build new ones. ERICA should assist in this effort, by also recognizing the need to build trust among strangers that might end up in the same team. Diversity here is also important, since artists want to work with like-minded people but also need the unexpected to occur. Again, it is a matter of balancing the two needs. The platform should further support scaffolding and incremental work, allowing groups of strangers to start from smaller projects that could escalate to something bigger if the parties agree. The opportunity to share the smaller or in-progress work for peer review and its potential marketability could solicit introspection on the quality of both the product and the group. Motivation schemes and transparency are important when people are involved in online work. Since technology changes the nature of (creative) work, it is essential to explore new ownership, labor and governance models that enhance creativity. Finally, ERICA should allow artists to maintain sustainability of their work and provide the means for opening to the market and efficient distribution of the creative outcomes.


\end{document}